# Seer: Leveraging Big Data to Navigate The Increasing Complexity of Cloud Debugging

Yu Gan
*Cornell University*

Meghna Pancholi
*Cornell University*

Dailun Cheng
*Cornell University*

Siyuan Hu
*Cornell University*

Yuan He
*Cornell University*

Christina Delimitrou
*Cornell University*

## Abstract

Performance unpredictability in cloud services leads to poor user experience, degraded availability, and has revenue ramifications. Detecting performance degradation a posteriori helps the system take corrective action, but does not avoid the QoS violations. Detecting QoS violations after the fact is even more detrimental when a service consists of hundreds of thousands of loosely-coupled microservices, since performance hiccups can quickly propagate across the dependency graph of microservices. In this work we focus on anticipating QoS violations in cloud settings to mitigate performance unpredictability to begin with. We propose *Seer*, a cloud runtime that leverages the massive amount of tracing data cloud systems collect over time and a set of practical learning techniques to signal upcoming QoS violations, as well as identify the microservice(s) causing them. Once an imminent QoS violation is detected Seer uses machine-level hardware events to determine the cause of the QoS violation, and adjusts the resource allocations to prevent it. In local clusters with 10 40-core servers and 200-instance clusters on GCE running diverse cloud microservices, we show that Seer correctly anticipates QoS violations 91% of the time, and attributes the violation to the correct microservice in 89% of cases. Finally, Seer detects QoS violations early enough for a corrective action to almost always be applied successfully.

## 1 Introduction

Cloud computing services are governed by strict quality of service (QoS) constraints in terms of throughput, and more critically tail latency [10,14,16,18]. Violating these requirements worsens the end user experience, leads to loss of availability and reliability, and has severe revenue implications [9,10,14,20,21]. A recent shift from monolithic designs to loosely-coupled microservices is aimed at improving service deployment, isolation, and modularity, but at the same time puts more pressure on performance predictability, as the latency requirements of each individual microservice is often in the microsecond granularity. Similarly, as datacenter servers become increasingly heterogeneous with the addition of FPGAs, hardware accelerators, and network offload engines, performance predictability becomes even more challenging.

The need for performance predictability has prompted a long line of work on performance tracing, monitoring, and debugging systems [13, 25, 31, 32, 35, 36]. Systems like Dapper and GWP rely on distributed tracing (often at RPC level) to detect performance abnormalities, while the Mystery Machine [13] leverages the large amount of logged data to extract the causal relationships between messages, and sidestep the challenge of clock synchronization across large clusters. This dependency model between requests can then be used towards performance optimizations, such as incremental result propagation that leverages the latency slack of certain requests.

In this work we take performance debugging one step further. Specifically, we note that detecting QoS violations after the fact, although useful to amend prolonged degraded performance caused by events like misconfigurations, machine failures, etc. still incurs the poor user experience and revenue implications discussed above. Even more, the longer the system operates under degraded performance the longer it takes for corrective actions to take effect and restore nominal performance. In settings where clouds applications are comprised of microservices this introduces the additional challenge of having to pinpoint the culprit of a QoS violation, a non-trivial task given that dependencies between microservices can cause unpredictable performance to propagate and amplify through the system.

Given the consequences of a posteriori QoS violation detection, we set out to answer the following question: *(i) can QoS violations be anticipated in cloud systems that host microservices, and (ii) can we pinpoint which microservice is the culprit of an upcoming QoS violation*

*early enough to take corrective action?*

Initially, anticipating performance degradations seems infeasible given that the vast majority of QoS violations are caused by unpredictable, short-term transient effects [33]. An aid in this attempt is the massive amount of data cloud systems collect about the execution of services they host over time. By mining this information in a practical, online manner we can detect QoS violations just early enough to avoid them altogether by taking actions, such as adjusting resource allocations.

We present *Seer*, a cloud monitoring and performance debugging system that leverages deep learning to diagnose and prevent QoS violations in a practical and online manner. The neural network in Seer is trained offline on RPC-level annotated execution traces collected using Apache Thrift's timing interface [1]. When the system is online, Seer takes streaming execution traces as input, and outputs the microservice (if any) that will cause a QoS violation in the near future. The execution traces capture the queue depth in front of each microservice at fine-grain intervals; we also experimented with latency and utilization traces and show that unlike queue depths they do not correlate as closely with performance. While Seer converges quickly for small clusters, as systems scale convergence time does to. To speedup inference at scale, we offloaded the neural network computation on Google's TPU public cloud offering, which yielded almost two orders of magnitude faster convergence. The current design converges within 5-10msec for a neural network with several hundred input and output neurons, and 5 hidden layers.

We evaluate Seer both in our local cluster of 10 40-core machines, and on a 200-instance cluster on Google Compute Engine. In our local cluster Seer correctly identifies upcoming QoS violations in 93% of cases, and correctly pinpoints the microservice initiating the QoS violation 91% of the time. In the GCE cluster, it correctly detects QoS violations 91% of the time, and correctly identifies the culprit in 89% of cases. For the cases where QoS violations are anticipated correctly, Seer is able to adjust resource allocations to prevent them altogether in most cases. We believe that systems like Seer that take a data-driven approach can make the management of complex cloud systems more practical, and we are currently working to make the system more scalable and robust to server heterogeneity, missing or noisy input traces, and techniques like autoscaling.

## 2 The Design of Seer

### 2.1 Distributed Tracing

A major challenge with microservices is that one cannot simply rely on the client to report performance as with traditional client-server applications. We developed a distributed tracing system that records latencies at RPC granularity using the Thrift timing interface. RPCs and REST requests are timestamped upon arrival and departure from each microservice by the tracing module, and data is aggregated in a centralized Cassandra database. The design of the tracing system is similar to Zipkin [8]. We additionally track the number of requests queued in each microservice, and distinguish between the time spent processing network requests and the time that goes towards application computation. In all cases the overhead from tracing is negligible, less than 0.1% on end-to-end latency, which is tolerable for such systems [13, 31, 32].

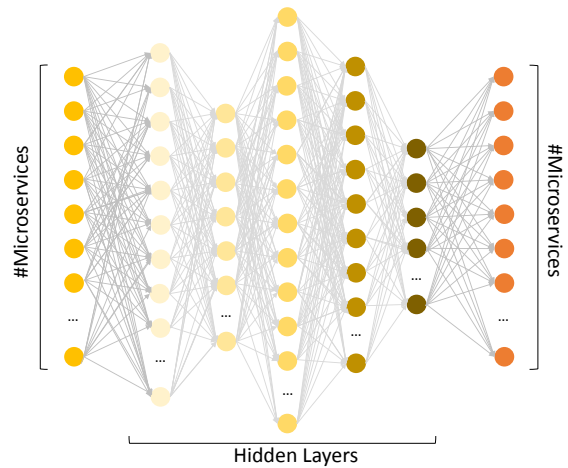

Figure 1: The neural network design in Seer.

### 2.2 Learning-Driven Performance Debugging

The key idea in Seer is that conditions that led to QoS violations in the past can be used to predict QoS violations in the near future. A popular way to model performance in cloud systems, especially when there are dependencies between tasks, is *queueing networks*. Although queueing networks are a valuable tool to model how bottlenecks propagate through the system, they do not capture all sources of contention in real systems, including the operating system and the network stack, and are therefore not accurate enough to predict upcoming QoS violations.

Instead in Seer, we take a data-driven approach that assumes no a priori knowledge on the architecture of a service, making the system robust to changing and unknown applications. A deep neural network is trained on the distributed traces collected using the system above to anticipate future QoS violations based on past system behavior. There are two factors that determine the accuracy a neural network can achieve. First, the metric that is used as input, and second, the configuration of the network's neurons and layers. We experimented with resource utilization, latency, and queue depths as input metrics. Consistent with prior work, utilization was not a good proxy for unpredictable performance [16, 17, 20, 28, 29]. Similarly latency (or the rate of latency increase) led to a large number of false positives, or signaled the wrong microservice as the QoS violation culprit. False negatives corresponded primarily to computationally-intensive applications that incur higher latencies than the rest of microservices, while signaling the wrong microservice as the one responsible for the QoS violation corresponded to microservices connected via a blocking dependency with services that were the real culprits. Again consistent with queueing theory [27] and prior work [19, 22, 26, 28], per-microservice queue depths consistently captured performance bottlenecks and pinpointed the microservices causing them.

| Service | Comm Protocol | Per-language LoC breakdown of end-to-end service |
|---|---|---|
| **Social network** | **RPC** | 34% C, 23% C++, 18% Java, 7% node.js, 6% Python, 5% Scala, 3% PHP, 2% Javascript, 2% Go |
| **Movie streaming** | **RPC** | 30% C, 21% C++, 20% Java, 10% PHP, 8% Scala, 5% node.js, 3% Python, 3% Javascript |
| **E-commerce** | **REST** | 21% Java, 16% C++, 15% C, 14% Go, 10% Javascript, 7% node, 5% C#, 5% Scala, 4% HTML, 3% Ruby |

Table 1: Code composition of each end-to-end service.

The second challenge, tuning the configuration parameters in the network (learning rate *a*, hidden layers, batch size, hidden units per layer) is done empirically. Figure 1 shows the neural network in Seer. The number of input and output neurons is equal to the number of active microservices in the cluster, with each input neuron capturing the queue depth of the corresponding microservice, and each output neuron firing if/when that microservice is about to initiate a QoS violation in the near future. All microservices in our setting run in single-constrained Docker containers, i.e., only a single microservice runs per container. This simplifies scaling up/out individual microservices independently. In Section 5 we discuss the implications of the number of active microservices changing as a result of techniques like autoscaling. To configure the network, we keep the number of input and output neurons constant, and first configure the learning rate *a* using ADAGRAD [23]. We then explore the impact of the number of hidden layers and units per hidden layer on output quality. The five hidden layers shown in Fig. 1 maximize the detection accuracy across a diverse testing set of application and system configurations, disjoint from the trace set the network is trained on (see the Validation section below). Weights and biases are obtained via Stochastic Gradient Descent (SGD) [11, 34].

**Training process:** Seer is trained on execution traces collected from all active microservices over time. Training happens offline, and only needs to be repeated when the server configurations or the type of active microservices change substantially. Traces from multiple servers are synchronized, and include requests queued per microservice over time. Training traces include annotated QoS violations; for now annotation is supervised manually, however we are exploring ways to completely automate the annotation process.

**Inference process:** During normal operation, execution traces are streamed through the network every few milliseconds and potential upcoming QoS violations are signaled. Once an imminent QoS violation is detected, Seer takes corrective action by first determining why the microservice is misbehaving, and then adjusting the resource allocation of the offending microservice to mitigate the unpredictable performance. In Section 4 we show an example of system evolution with and without Seer's intervention.

## 3 Validation

### 3.1 Methodology

**Applications:** Although there are many open-source microservices that could serve as individual components of an end-to-end service, there are no representative end-to-end applications built with microservices, with the exception of Sockshop, an e-commerce site by Weave [7]. To address this we have developed three end-to-end services which we plan to open-source, each consisting of tens of different microservices, and implementing a *social network*, a *movie streaming* service, and an *e-commence site* based on Sockshop. Individual microservices include `nginx` [5], `memcached` [24], `mongodb` [4], `RabbitMQ` [3], and `Apache http server`, among others. Table 1 shows a breakdown of each end-to-end service per language, which highlights the software heterogeneity that is often present in microservices. We additionally built an RPC framework over Apache Thrift [1] to connect individual microservices in the *social network* and *movie streaming* service. Microservices in the *e-commerce site* are connected over http.

**Systems:** We use two clusters settings. First, a dedicated local cluster with 10, 2-socket 40-core servers with 128GB of RAM each. Each server is connected to a 40Gbps ToR switch over 10Gbe NICs. Second, we use a 200-instance cluster on Google Compute Engine (GCE) to study the scalability of Seer. All instances are `n1-standard-64`, each with 64 vCPUs and 240GB of

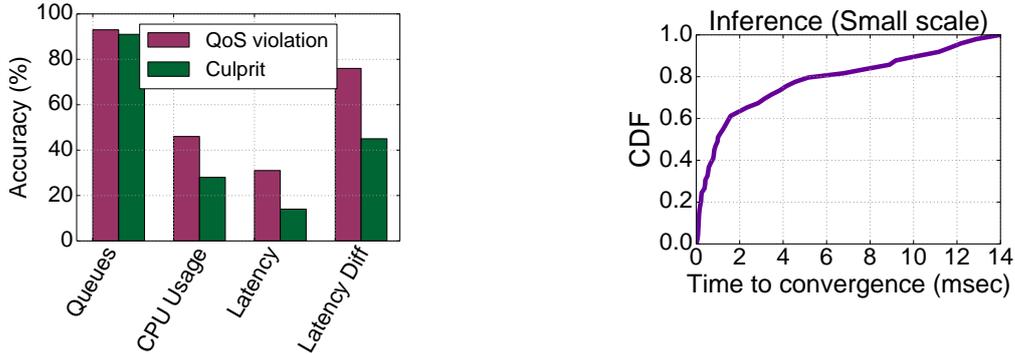

Figure 2: (a) The accuracy of detecting upcoming QoS violations when using different metrics as inputs of the neural network. CPU utilization and per-microservice latency are clearly not sufficient to capture events that lead to unpredictable performance. Using the rate with which per-microservice latency changes improves accuracy, however it still misses a significant fraction of QoS violations. The per-microservice queue depth captures almost all QoS violations, and for the majority of them pinpoints the correct microservice as the QoS violation culprit. (b) The time it takes for inference to converge in Seer, for the small-scale 10 server cluster.

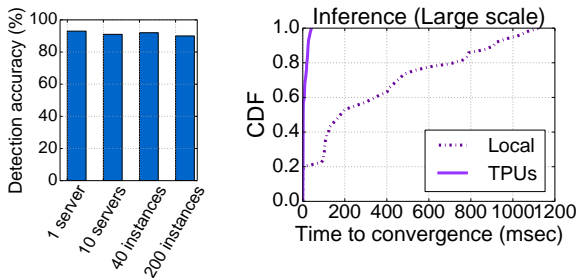

Figure 3: (a) The accuracy of QoS violation detection is Seer as the cluster size increases, and (b) The CDF of convergence time with our local implementation, and using TPUs on Google Cloud for the 200-instance cluster.

RAM.

## 3.2 Evaluation

**Accuracy:** Fig. 2a shows the detection accuracy in Seer under different input metrics. CPU utilization and per-microservice latencies miss the majority of QoS violations and mislabel the microservices initiating the violations. Using the rate at which per-microservice latency changes achieves higher accuracy but still misses a significant fraction of QoS violations, and incorrectly labels a disproportionate fraction of microservices as culprits. Using the per-microservice queue depth as the input of the neural network captures the majority of QoS violations, and pinpoints the responsible microservice.

**Convergence time:** Fig. 2b shows the convergence time for inference in the small 10-server cluster. For 60% of QoS violations, detection happens within 2msec from obtaining the per-microservice traces, early enough to apply most corrective actions that avoid the QoS violation altogether. Even the QoS violations in the high percentiles of the CDF are detected within 14msec at most, which is usually sufficient for the system to react.

**Scalability:** We now examine the accuracy and scalability of Seer as the size of the cluster increases. Fig. 3a shows QoS detection accuracy for different cluster sizes; the 1 and 10 server settings are local, while the 40- and 200-instance clusters are on GCE. The results show that Seer is robust to the size of the cluster in terms of detection accuracy. Fig. 3b however shows the penalty of scalability in convergence time for the 200-instance cluster. For the majority of cases it takes several hundreds of msec for the network to converge, at which point the QoS violation has occurred. To address this we rewrote Seer using Tensorflow and ported it on the Google TPU public cloud. The change in inference time is dramatic, two orders of magnitude in some cases, ensuring that detection happens early enough to be meaningful.

## 4 QoS Violation Prevention

Once an upcoming QoS violation is detected Seer takes action to try to avoid it. This involves first determining what will cause the QoS violation before it manifests as an increase in tail latency. To do so Seer looks at hardware-level per-resource utilization statistics on the machine where the offending microservice resides. This includes CPU and memory utilization, memory, network, and I/O bandwidth usage, and last level cache misses. Although this is not an exhaustive list of resources where contention can cause QoS violations, in practice it covers a large fraction of performance degradations.

Once the problematic resource is located, Seer adjusts the resource allocation, either resizing the Docker container, or using mechanisms like Intel's Cache Allocation Technology (CAT) for last level cache (LLC) partitioning, and the Linux traffic control's hierarchical token bucket (HTB) queueing discipline in qdisc [12, 30] for network bandwidth partitioning.

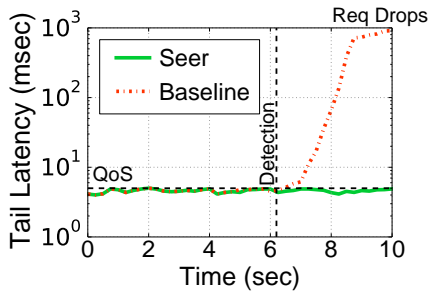

Figure 4: Tail latency with and without Seer. Once the upcoming QoS violation is detected Seer determines which resource is causing the unpredictable performance is the offending microservice and adjusts its allocation. Post detection, tail latency with Seer is nominal, while if the QoS violation remains undetected tail latency continues to worsen until requests start getting dropped.

Fig. 4 shows the impact on tail latency with and without Seer. Once the upcoming QoS violation is detected Seer determines the resource that will cause it, in this case insufficient LLC capacity, and uses CAT to adjust it. Post detection the service's tail latency with Seer remains nominal, while if the QoS violation had remained undetected tail latency would continue to worsen until requests started getting dropped. Note that once the system arrives in such a problematic state it, returning to normal operation has significant inertia.

Some of the resource measurements Seer uses for detect problematic resources involve access to hardware performance counters. Unfortunately public clouds do not enable access to such events. In that case, Seer uses a set of contentious microbenchmarks, each targeting a different system resource to pinpoint problematic resources [15]. For example, a cache thrashing microbenchmark will reveal cache saturation, while a network bandwidth-demanding microbenchmark will reveal insufficient bandwidth allocations. These microbenchmarks need to run for a couple milliseconds before signaling the resource under contention.

## 5 Discussion

Seer is early work, but is already used by several research groups at Cornell and elsewhere. Nonetheless, the present design has a number of limitations, which we are resolving in current work. First, because the number of input and output neurons in Seer is equal to the number of active microservices, the system is not robust to techniques like autoscale which spawn additional or terminate existing containers on the cluster. We are currently expanding the design first, to incrementally train the network under a varying number of active microservices, and second, to automatically derive the architecture of the end-to-end service, including the ordering of dependencies between microservices to make the system more robust to changes in the number of active microservices.

Second, Seer assumes full control over the entire cluster, or at least of individual servers, i.e., it assumes that it can collect traces from all active microservices on a physical machine. This may not always be the case, especially in public clouds, or when using third-party applications that cannot easily be instrumented. We are extending the system design to tolerate missing or noisy tracing information, which also addresses the issue of multiple microservices per container.

Finally, even though Seer is able to avert the majority of QoS violations, there are still some events that are not predicted early enough for corrective actions to take place. These typically involve memory-bound microservices, where the memory subsystem is saturated. Memory, like any storage medium, has inertia, so resource adjustment decisions require longer to take effect. We are exploring whether predicting further into the future is possible without significantly increasing the number of false positives, or whether alternative resource isolation mechanisms like cache partitioning can help alleviate memory pressure faster.

## 6 Future Work

Cloud systems and applications continuously increase in size and complexity. The recent switch from monoliths to microservices puts even more pressure on performance predictability, and at the same time makes manual performance debugging impractical. In this paper we presented early work on Seer, a monitoring and performance debugging runtime that leverages practical learning techniques and the massive amount of tracing data cloud systems collect to proactively detect QoS violations with enough slack to prevent them altogether. We have evaluated Seer both on local clusters and a large cluster on GCE and validated its accuracy in anticipating QoS violations and pinpointing the microservices that cause them. As cloud applications shift from batch to low-latency, systems like Seer can improve their QoS, predictability, and responsiveness in a practical, online manner. Finally, performance predictability is important not only in the cloud, but in IoT settings as well. Such environments are additionally plagued by unreliable communication channels, intermittent connectivity, limited battery lifetime, and constrained on-board resources. We are currently expanding Seer's applicability to swarms of heterogeneous edge devices, such as programmable drones [6] and Elisa-3 robots [2].

# References

[1] Apache thrift. https://thrift.apache.org.

[2] Elisa-3. http://www.gctronic.com/doc/index.php/Elisa-3.

[3] Messaging that just works. https://www.rabbitmq.com/.

[4] mongodb. https://www.mongodb.com.

[5] Nginx. https://nginx.org/en.

[6] Parrot ar.drone 2.0 edition. https://www.parrot.com/us/drones/parrot-ardrone-20-elite-edition.

[7] Sockshop: A microservices demo application. https://www.weave.works/blog/sock-shop-microservices-demo-application.

[8] Zipkin. http://zipkin.io.

[9] BARROSO, L. Warehouse-scale computing: Entering the teenage decade. *ISCA Keynote, SJ, June 2011*.

[10] BARROSO, L., AND HOELZLE, U. *The Datacenter as a Computer: An Introduction to the Design of Warehouse-Scale Machines*. MC Publishers, 2009.

[11] BOTTOU, L. Large-scale machine learning with stochastic gradient descent. In *Proceedings of the International Conference on Computational Statistics (COMPSTAT)*, Paris, France, 2010.

[12] BROWN, M. A. Traffic control howto. http://linux-ip.net/articles/Traffic-Control-HOWTO/.

[13] CHOW, M., MEISNER, D., FLINN, J., PEEK, D., AND WENISCH, T. F. The mystery machine: End-to-end performance analysis of large-scale internet services. In *Proceedings of the 11th USENIX Conference on Operating Systems Design and Implementation* (Berkeley, CA, USA, 2014), OSDI'14, USENIX Association, pp. 217–231.

[14] DEAN, J., AND BARROSO, L. A. The tail at scale. In *CACM, Vol. 56 No. 2*.

[15] DELIMITROU, C., AND KOZYRAKIS, C. iBench: Quantifying Interference for Datacenter Workloads. In *Proceedings of the 2013 IEEE International Symposium on Workload Characterization (IISWC)*, Portland, OR, September 2013.

[16] DELIMITROU, C., AND KOZYRAKIS, C. Paragon: QoS-Aware Scheduling for Heterogeneous Datacenters. In *Proceedings of the Eighteenth International Conference on Architectural Support for Programming Languages and Operating Systems (ASPLOS)*, Houston, TX, USA, 2013.

[17] DELIMITROU, C., AND KOZYRAKIS, C. QoS-Aware Scheduling in Heterogeneous Datacenters with Paragon. In *ACM Transactions on Computer Systems (TOCS), Vol. 31 Issue 4*, December 2013.

[18] DELIMITROU, C., AND KOZYRAKIS, C. Quality-of-Service-Aware Scheduling in Heterogeneous Datacenters with Paragon. In *IEEE Micro Special Issue on Top Picks from the Computer Architecture Conferences*, May/June 2014.

[19] DELIMITROU, C., AND KOZYRAKIS, C. Quasar: Resource-Efficient and QoS-Aware Cluster Management. In *Proc. of ASPLOS*, Salt Lake City, 2014.

[20] DELIMITROU, C., AND KOZYRAKIS, C. HCloud: Resource-Efficient Provisioning in Shared Cloud Systems. In *Proceedings of the Twenty First International Conference on Architectural Support for Programming Languages and Operating Systems (ASPLOS)* (April 2016).

[21] DELIMITROU, C., AND KOZYRAKIS, C. Bolt: I Know What You Did Last Summer... In The Cloud. In *Proc. of the Twenty Second International Conference on Architectural Support for Programming Languages and Operating Systems (ASPLOS)* (2017).

[22] DELIMITROU, C., SANCHEZ, D., AND KOZYRAKIS, C. Tarcil: Reconciling Scheduling Speed and Quality in Large Shared Clusters. In *Proceedings of the Sixth ACM Symposium on Cloud Computing (SOCC)* (August 2015).

[23] DUCHI, J., HAZAN, E., AND SINGER, Y. Adaptive subgradient methods for online learning and stochastic optimization. In *Journal of Machine Learning Research 12 (2011) 2121-2159*, 2011.

[24] FITZPATRICK, B. Distributed caching with memcached. In *Linux Journal, Volume 2004, Issue 124, 2004*.

[25] FONSECA, R., PORTER, G., KATZ, R. H., SHENKER, S., AND STOICA, I. X-trace: A pervasive network tracing framework. In *Proceedings of the 4th USENIX Conference on Networked Systems Design & Implementation* (Berkeley, CA, USA, 2007), NSDI'07, USENIX Association, pp. 20–20.

[26] GENG, Y., LIU, S., YIN, Z., NAIK, A., PRABHAKAR, B., ROSENBLUM, M., AND VAHDAT, A. Exploiting a natural network effect for scalable, fine-grained clock synchronization. In *Proc. of NSDI*, 2018.

[27] GROSS, D., SHORTLE, J. F., THOMPSON, J. M., AND HARRIS, C. M. Fundamentals of queueing theory. In *Wiley Series in Probability and Statistics, Book 627*, 2011.

[28] KASTURE, H., AND SANCHEZ, D. Ubik: Efficient Cache Sharing with Strict QoS for Latency-Critical Workloads. In *Proceedings of the 19th international conference on Architectural Support for Programming Languages and Operating Systems (ASPLOS-XIX)* (March 2014).

[29] LO, D., CHENG, L., GOVINDARAJU, R., BARROSO, L. A., AND KOZYRAKIS, C. Towards energy proportionality for large-scale latency-critical workloads. In *Proceedings of the 41st Annual International Symposium on Computer Architecuture (ISCA)*, Minneapolis, MN, 2014.

[30] LO, D., CHENG, L., GOVINDARAJU, R., RANGANATHAN, P., AND KOZYRAKIS, C. Heracles: Improving resource efficiency at scale. In *Proc. of the 42Nd Annual International Symposium on Computer Architecture (ISCA)*, Portland, OR, 2015.

[31] REN, G., TUNE, E., MOSELEY, T., SHI, Y., RUS, S., AND HUNDT, R. Google-wide profiling: A continuous profiling infrastructure for data centers. *IEEE Micro* (2010), 65–79.

[32] SIGELMAN, B. H., BARROSO, L. A., BURROWS, M., STEPHENSON, P., PLAKAL, M., BEAVER, D., JASPAN, S., AND SHANBHAG, C. Dapper, a large-scale distributed systems tracing infrastructure. Tech. rep., Google, Inc., 2010.

[33] SINGH, A., ONG, J., AGARWAL, A., ANDERSON, G., ARMISTEAD, A., BANNON, R., BOVING, S., DESAI, G., FELDERMAN, B., GERMANO, P., KANAGALA, A., PROVOST, J., SIMMONS, J., TANDA, E., WANDERER, J., HLZLE, U., STUART, S., AND VAHDAT, A. Jupiter rising: A decade of clos topologies and centralized control in googles datacenter network. In *Sigcomm '15* (2015).

[34] WITTEN, I. H., FRANK, E., AND HOLMES, G. *Data Mining: Practical Machine Learning Tools and Techniques*. 3rd Edition.

[35] XU, T., JIN, X., HUANG, P., ZHOU, Y., LU, S., JIN, L., AND PASUPATHY, S. Early detection of configuration errors to reduce failure damage. In *Proceedings of the 12th USENIX Conference on Operating Systems Design and Implementation* (Berkeley, CA, USA, 2016), OSDI'16, USENIX Association, pp. 619–634.

[36] YU, X., JOSHI, P., XU, J., JIN, G., ZHANG, H., AND JIANG, G. Cloudseer: Workflow monitoring of cloud infrastructures via interleaved logs. In *Proceedings of the Twenty-First International Conference on Architectural Support for Programming Languages and Operating Systems* (New York, NY, USA, 2016), ASPLOS '16, ACM, pp. 489–502.